\documentclass[twocolumn,showpacs,floats,floatfix,superscriptaddress,aps,pra]{revtex4-1}
\usepackage{amsfonts}
\usepackage{amssymb}
\usepackage{amsmath}
\usepackage{calc}
\usepackage{graphicx}
\usepackage{bm}

\usepackage[normalem]{ulem}
\usepackage{amsmath,amssymb}
\usepackage{multirow}
\usepackage{xcolor,soul}
\usepackage{srcltx}
\usepackage{hyperref,graphicx}

\def\be{ \begin{equation}}
\def\ee{ \end{equation}}
\def\bea{ \begin{eqnarray}}
\def\eea{ \end{eqnarray}}
\def\bse{ \begin{subequations}}
\def\ese{ \end{subequations}}
\def\bc{ \begin{center}}
\def\ec{ \end{center}}

\def\sech{\,\text{sech}}

\begin{document}

\author{Stefano Longhi}
\affiliation{Dipartimento di Fisica, Politecnico di Milano and Istituto di Fotonica e Nanotecnologie del Consiglio Nazionale delle Ricerche, Piazza L. da Vinci 32, I-20133 Milano, Italy}
\author{Giuseppe Della Valle}
\affiliation{Dipartimento di Fisica, Politecnico di Milano and Istituto di Fotonica e Nanotecnologie del Consiglio Nazionale delle Ricerche, Piazza L. da Vinci 32, I-20133 Milano, Italy}

\title{Absence of Floquet scattering in oscillating non-Hermitian potential wells}
\date{\today }

\begin{abstract}
Scattering of a quantum particle from an oscillating 
barrier or well does not generally conserve the particle energy owing to energy exchange with the photon field, and an incoming particle-free state is scattered into
a set of outgoing (transmitted and reflected) free states according to Floquet scattering theory. 
Here we introduce two families of oscillating non-Hermitian potential wells in which Floquet scattering is fully suppressed for any energy of the incident particle. 
The scattering-free oscillating potentials are synthesized by application of the Darboux transformation to the time-dependent Schr\"{o}dinger equation. For one of the two families of scattering-free potentials, the oscillating potential turns out to be fully invisible. 
\end{abstract}

\pacs{
03.65.Nk  ,	
03.65.Xp  ,	
42.25.Fx  ,	
11.30.Er 	
 }
\maketitle

\section{Introduction}

Scattering dynamics from periodic time-varying
potentials is of fundamental importance in different areas of physics, involving
fundamental aspects of quantum mechanics, such as the problem of tunneling times \cite{Butt1,Butt2,Butt3}, classical and quantum chaos \cite{chaos1,chaos2,chaos3,chaos4,book}, 
and the quantum transport properties of microscopic and mesoscopic
systems \cite{Platero,Hanggi,tras1,tras2,tras3,meso}.  
Several studies have investigated in details the scattering properties of time-varying potential barriers or wells based on the Floquet formalism (see, e.g., \cite{Flo}), with applications to problems such as photon-assisted tunneling \cite{Platero,r3,r4},
quantum pumping \cite{r5c,r5b,r5,r5graphene}, electron scattering by intense
laser-driven potentials \cite{r6}, and electron transport in graphene with modulated barriers \cite{tras2,tras3,r5graphene}. These systems can display
a quite rich quantum and classical dynamics, such as  chaotic
scattering and chaos-assisted tunneling \cite{r7,r8,r9,r10,r11}, coherent destruction of
tunneling \cite{r12,Grifoni}, quantum interference \cite{r13}, Fano resonances \cite{tras2,Flo}, field-induced barrier transparency \cite{Vor}, and particle-field entanglement in the second-quantization regime \cite{quantum}. 
Noticeably, the scattering dynamics from a periodic time-varying
potential barrier or well is analogous, under certain conditions, to Bragg scattering of monochromatic matter or classical waves 
from a periodic (grating) potential \cite{Berry1,Berry2,Longhi}. For example, a quantum-optical analogy can be established between the well-known problem of Bragg scattering of light waves from a diffraction grating under grazing incidence and the  scattering dynamics of a non-relativistic quantum particle from a periodic time-varying
potential well or barrier \cite{book,Longhi}. \par 
In recent years,  a great and increasing interest has been devoted to study the properties of quantum systems described by a {\it non-Hermitian} Hamiltonian \cite{Moi}, especially in the context $\mathcal{PT}$-symmetric quantum mechanics \cite{Bender}.  Scattering in non-Hermitian potentials has been experimentally  investigated, for example, in the diffraction of matter waves from complex optical potentials \cite{Keller,Ober} and for light transport  in optical structures with tailored gain and loss regions \cite{cazz1,cazz2,cazz3}. 
 Non-Hermiticity has been shown to strongly affect Bragg scattering in imaginary periodic potentials, with important effects such as the violation of the Friedel's law \cite{Berry1,Keller} and unidirectional invisibility \cite{Lin,Longhi10,Longhi11,Jones}. 
The scattering properties of static and non-Hermitian potential wells and barriers have been investigated in several previous works  as well \cite{NH1,NH2,NH3,NH4,NH5}. In particular, it was shown rather generally that, as opposed to an Hermitian potential barrier or well, the reflectance in a non-Hermitian  potential barrier is generally different for left and right incidence \cite{NH1,NH3}.  However, the scattering properties of 
imaginary potential barriers or wells periodically oscillating in time have not received attention to date. \par
In this work we consider the scattering properties from exactly-solvable time-periodic complex potential wells, which show the rather exceptional property to appear {\it reflectionless} and to {\it conserve} the energy of the transmitted particle at any energy of the incident particle \cite{note1}. This means that the particle, in addition to not being  reflected from the potential well, does not exchange energy with the photon field, regardless the initial energy of the particle, i.e. the scattering matrix elements to all the Floquet channels \cite{Flo} vanish. In the analogous problem of wave diffraction from a complex grating potential \cite{Berry1,Berry2,Longhi} discussed above, this means the entire absence of grating diffraction orders, both in reflection and transmission \cite{note2}, which are analogous to the Floquet scattering channels in the time-periodic case. We note that, as the existence of {\it static}  reflectionless potentials is well-known in quantum physics (see, e.g., \cite{uff1,uff2,uff3,uff4}), including the case of non-Hermitian potential wells \cite{NH4}, the existence of reflectionless and energy-conserving potentials {\it periodically-oscillating in time} found in our work has not been predicted yet to the best of our knowledge and provides a nontrivial result in the basic scattering theory of oscillating potentials. \par
The paper is organized as follows. In Sec.II the basic principles of particle scattering form an oscillating potential are briefly reviewed, and its connection to Bragg scattering of matter or classical waves from a diffraction grating is discussed. In Sec.III two exactly-solvable families of scattering-free non-Hermtian oscillating potentials are introduced by application of the Darboux transformation for the time-dependent Scrh\"{o}dinger equation. The absence of Floquet scattering is checked by numerical simulations of the Schr\"{o}dinger equation in Sec.IV. Finally, the main conclusions and a brief discussion of open questions are presented in Sec.V.

\section{Floquet scattering from a time-periodic potential}
The scattering of a quantum particle from an oscillating potential is described by the time-dependent Schr\"{o}dinger equation (with $\hbar=m=1$)
\begin{equation}
i \frac{\partial \psi}{\partial t}= -\frac{1}{2} \frac{\partial^2 \psi}{\partial x^2}+V(x,t) \psi(x)
\end{equation}
where $V(x,t+T)=V(x,t)$ is the time-periodic potential with period $T= 2 \pi / \omega$.  We assume that the scattering potential is localized at around $x=0$ with  $V(x,t) \rightarrow 0$ as $ x \rightarrow \pm \infty$ [see Fig.1(a)], so that far from $x=0$ the asymptotic solutions to Eqs.(1) are  free-particle (plane wave) states.  It is worth mentioning that the scattering of a quantum particle from an oscillating potential with limited support, as described by Eq.(1), is analogous the the Bragg scattering of a monochromatic matter or optical wave from a two-dimensional diffraction grating, as shown in Fig.1(b) (see, for instance, \cite{Berry1,Berry2,Longhi}). For example, let us consider the scattering of a monochromatic light wave at wavelength (in vacuum) $\lambda_0$ from a two-dimensional dielectric grating with refractive index $n(x,z)=n_0-\Delta n(x,z)$, where $\Delta n \ll n_0$ is the index change from the substrate refractive index $n_0$ and the grating potential $\Delta n(x,z)$ is periodic, along the spatial direction $z$, with period $\Lambda \ll \lambda_0$, i.e. $\Delta n(x,z +\Lambda)=\Delta n(x,z)$. For a TE-polarized optical wave ($E_x=E_z=H_y=0$), the electric field $E_y(x,z)$ satisfies the Helmholtz equation $(\partial^2_x+\partial_z^2) E_y+k_0^2n^2(x,z)E_y=0$, where $k_0=2 \pi / \lambda_0$. For grazing incidence, after setting $E_y(x,z)= \psi(x,z) \exp(i k_0n_0z)$ the field envelope $\psi(x,z)$ satisfies the Schr\"{o}dinger-like (paraxial) wave equation \cite{Longhi}
\begin{equation}
i  \frac{\partial \psi}{\partial z}=-\frac{1}{2n_0k_0} \frac{\partial^2 \psi}{\partial x^2}+k_0 \Delta n(x,z) \psi
\end{equation}
which has the same form as Eq.(1), provided that the temporal variable $t$ is replaced by the spatial (paraxial) propagation distance $z$ and the potential $V(x,t)$ is defined by the index change $\Delta n(x,z)$. In the following, we will specifically refer to the scattering problem of a quantum particle from an oscillating potential in the framework of Eq.(1), however the same analysis holds for the Bragg scattering problem at grazing incidence on a diffraction grating.\par

\begin{figure}[t]
\includegraphics[width=8cm]{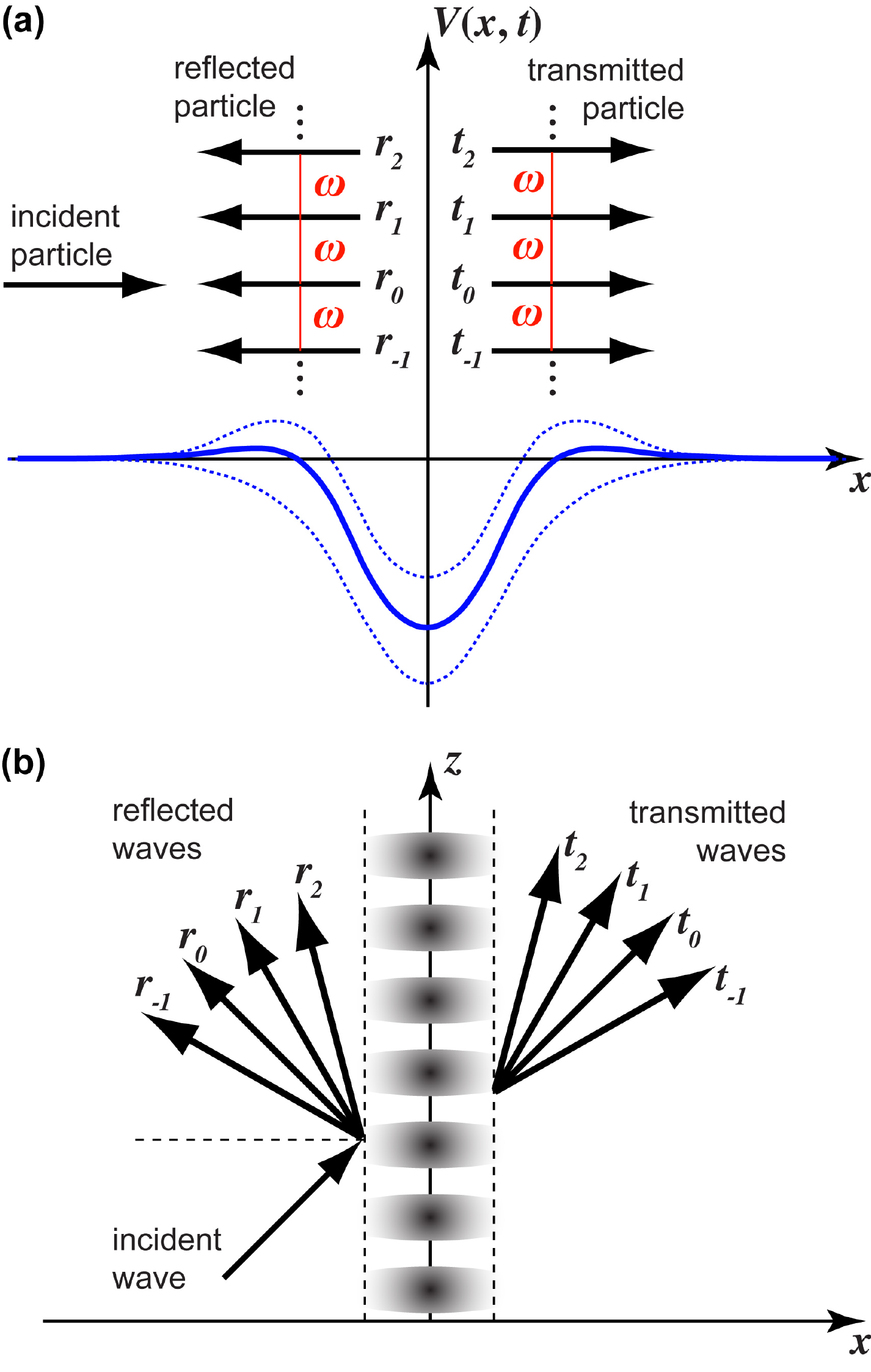}
\caption{(Color online). (a) Schematic of Floquet scattering of a quantum particle from a potential oscillating in time at frequency $\omega$. The energy of the incidence particle is $E$. The reflected and transmitted particle states correspond to the various Floquet  channels at energies $E+n \omega$. (b) Schematic of Bragg diffraction of a light or matter wave from a grating potential at grazing incidence. The reflected and transmitted diffraction orders from the grating correspond to the Floquet channels in (a).}
\end{figure}

  Let us consider a particle with momentum $p$ and energy $E=p^2/2$ incident onto the scattering potential from the left side. The solution to Eq.(1) has the form 
  \begin{equation}
  \psi(x,t)=\psi_{in}(x,t)+\psi_{sc}(x,t),
  \end{equation}
   where 
   \begin{equation}
   \psi_{in}(x,t)= \exp(ipx-Et)
   \end{equation}
    is the incident particle state and $\psi_{sc}(x,t)$ is the scattered state. According to Floquet theory, $\psi_{sc}(x,t)$ has the following asymptotic form
\begin{equation}
\psi_{sc}(x,t)= \sum_{n=-\infty}^{\infty} t_n^{(left)}(E) \exp(ip_nx-E_n t) 
\end{equation}
 for $x \rightarrow + \infty$, and 
\begin{equation}
\psi_{sc}(x,t)= \sum_{n=-\infty}^{\infty}  r_n^{(left)}(E) \exp(-ip_nx-E_n t) 
\end{equation}
 for $ x \rightarrow -\infty$, where $E_n=E+n \omega$, $p_n= \sqrt{2 (E+n \omega)}$ and ${\rm Im}(p_n) \geq 0$. The complex coefficients $t_n^{(left)}(E)$ and $r_n^{(left)}(E)$ for the propagative modes (i.e. for indices $n$ with $E+n \omega>0$, corresponding to a real-valued momentum $p_n$) define the transmission and reflection coefficients for the various Floquet channels at energies $E+ n \omega$ for left-side incidence \cite{Flo}. In a similar way, one can introduce the Floquet reflection and transmission coefficients 
 $t_n^{(right)}(E)$ and $r_n^{(right)}(E)$ for right-side particle incidence, which are formally obtained from Eq.(4) by the change $p \rightarrow -p$ and by reversing $+\infty$ and $-\infty$ in the asymptotic behavior Eqs.(5) and (6). Note that, as compared to the scattering from a static potential, in addition to a non-vanishing probability of the particle to be reflected from the potential, the particle energy $E$ is not conserved during the interaction because of energy exchange (absorption and emission) with the photon field. In the analogous problem of Bragg scattering from a diffraction grating [Fig.1(b)], the reflection and transmission coefficients of the various Floquet channels correspond to the reflection and transmission amplitudes of the various diffraction orders.

 \section{Oscillating quantum wells without Floquet scattering}
 Absence of Floquet scattering from an oscillating potential for left-side particle incidence can be defined by the conditions $r_n^{(left)}(E) \equiv 0$ for any $n$  with $E_n=E+n \omega >0$,  and $t_n^{(left)} (E) \equiv 0$ for $n \neq 0$, which imply the absence of reflected waves and the conservation of the energy for the transmitted particle.  Such  conditions extend to the time-periodic case the well-known case of {\it reflectionless} static potentials (see, e.g., \cite{uff1,uff2,uff3,uff4}), where only the reflection and transmission coefficients $r_0^{(left)}(E)$ and $t_0^{(left)}(E)$ are involved. Our goal is to synthesize an oscillating potential well which does not show Floquet scattering.  To this aim, we use the method of the Darboux transformation, well-known in problems of reflectionless static potentials, extended to the time-dependent Schr\"{o}dinger equation \cite{D1,D2}. Let us indicate by $\hat{\mathcal{H}}_0=-(1/2) \partial^2_x+V_0(x,t)$ and $\hat{\mathcal{H}}=-(1/2) \partial^2_x+V(x,t)$ two time-dependent Hamiltonians with potentials $V_0(x,t)$ and $V(x,t)$, respectively. A linear differential operator $\hat{\mathcal{D}}$ of first order in $x$, with time-dependent coefficients and satisfying the intertwining relation
 \begin{equation}
 \hat{\mathcal{D}} (i \partial_t-\hat{\mathcal{H}}_0)=(i \partial_t-\hat{\mathcal{H}}) \hat{\mathcal{D}}
 \end{equation}
 is called a first-order Darboux transformation operator. Its explicit form is given by \cite{D1,D2}
 \begin{equation}
 \hat{\mathcal{D}}=L(t) \left( \partial_x-\frac{1}{u} \frac{\partial u}{\partial x} \right)
 \end{equation}
 where $L(t)$ is an arbitrary function of time $t$ and $u(x,t)$ is any solution to the equation $(i \partial_t-\hat{\mathcal{H}}_0)u=0$. The intertwining relation (7) holds provided that the two potentials $V_0(x,t)$ and $V(x,t)$ are connected by the relation
 \begin{equation}
 V(x,t)=V_0(x,t)-\frac{\partial^2 \left( {\rm log} \; u \right)}{\partial x^2}-i \frac{1}{L}\frac{dL}{dt}.
 \end{equation}
 Note that, if $\phi(x,t)$ is any solution to the time-dependent Schr\"{o}dinger equation $i \partial_t \phi=\hat{\mathcal{H}}_0 \phi$, then the Darboux transform of $\phi(x,t)$, defined by
 \begin{equation}
 \psi(x,t)= \hat{\mathcal{D}} \phi(x,t)=L(t) \frac{\partial \phi}{\partial x}-\frac{L(t)}{u} \frac{\partial u}{\partial x} \phi(x,t)
 \end{equation}
 is a solution to the time-dependent Schr\"{o}dinger equation  $i \partial_t \psi=\hat{\mathcal{H}} \psi$. This result readily follows from the intertwining relation (7). Note that, for an initial real-valued potential $V_0(x,t)$, the partner potential $V(x,t)$ obtained from Eq.(9) turns out to be real-valued provided that the function $u(x,t)=|u(x,t)| \exp[i \theta(x,t)]$ has a phase $\theta(x,t)$ which is at most a quadratic function of $x$, i.e. $\partial_{xxx} \theta=0$; only under such a constraint, after setting $L(t)= \exp[- \int_0^t dt' \theta_{xx} (t')]$ the partner potential $V(x,t)$ turns out to be real-valued \cite{D1,D2}.  This is a rather severe constraint, and indeed the method of Darboux transformation to the time-dependent Schr\"{o}dinger equation in the Hermitian case has been applied to rather few and special cases \cite{D1,D2,D3,D4}.  In particular, the Hermitian constraint does not allow one to synthesize an oscillating quantum well with suppressed Floquet scattering, like for a static potential well (see, for instance, \cite{uff2}). Nevertheless,  if we remove the Hermiticity constraint, allowing the potential to become imaginary, an oscillating quantum well with suppressed Floquet scattering can be synthesized by the method of the Darboux transformation. \par
 \begin{figure}[t]
\includegraphics[width=8cm]{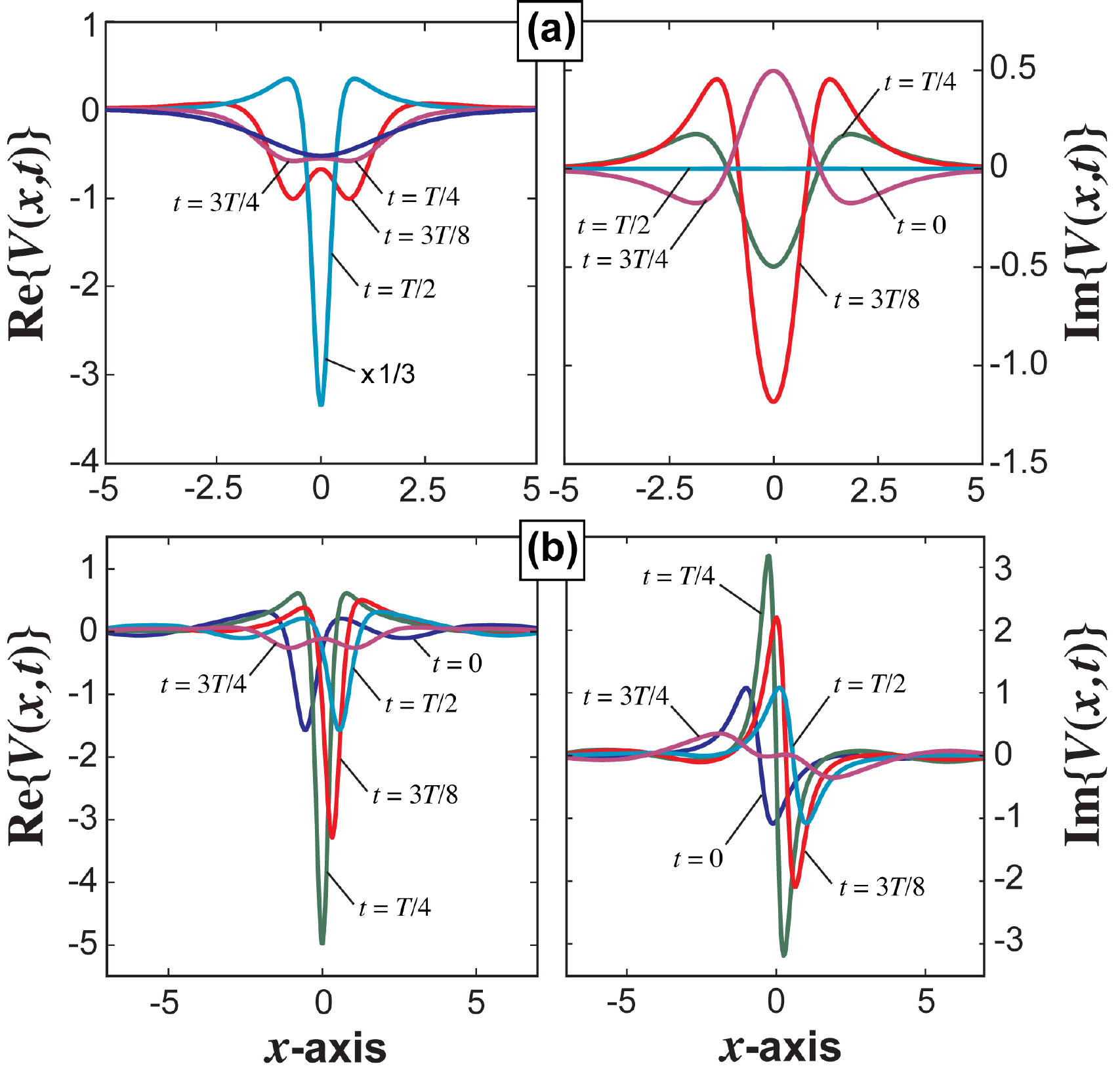}
\caption{(Color online). (a) Behavior of the real and imaginary parts of the potential (12) for $\alpha=0.9$, $\beta=0$, $\mu=1$ at times $t=0$, $t=T/4$, $t=3T/8$, $t=T/2$, and $t= 3T/4$. (b) Same as (a) but for the potential (19) [$\alpha=\beta=2$, $\mu=1$].}
\end{figure}

  To this aim, let us assume $V_0(x,t)=0$, and let us consider the Darboux operator $\mathcal{D}$ obtained by taking $L(t)=1$ in Eq.(8). We will consider specifically two distinct solvable cases. In the former case the Floquet scattering is fully suppressed, however the potential well introduces an advancement  of the transmitted particle, i.e. there is a non-vanishing energy-dependent phase of the transmission $t_0(E)$ like in the Hermitian reflectionless potential well \cite{uff2}.  
 In the latter case the Floquet scattering is fully suppressed and the oscillating potential well appears to be fully invisible, i.e. $t_0(E)=1$.\par
 The former case is obtained by assuming  the following solution  $u(x,t)$ to the equation $( i \partial_t-\hat{\mathcal{H}_0}) u=0$
 \begin{equation}
 u(x,t)=\alpha+ \beta x+ \cosh(\mu x) \exp(i \omega t)
 \end{equation}
 where $\mu>0$ is an arbitrary real-valued parameter, $\omega= \mu^2 /2$, and $\alpha$, $\beta$ are two arbitrary complex-valued parameters, satisfying the constraint $|\alpha+\beta x |< \cosh (\mu x)$ for $-\infty < x < \infty$. For example, by taking $\beta=0$ such a condition is satisfied provided that $|\alpha|<1$. This constraint  ensures that $u(x,t)$ does not vanish for any time $t$, thus avoiding singularities in the partner potential $V(x,t)$. Using Eqs.(9) and (11), the partner potential $V(x,t)$ associated to $V_0(x,t)=0$ via the Darboux transformation can be readily calculated and reads explicitly
\begin{eqnarray}
V(x,t)& = & - \frac{\mu^2 \cosh (\mu x) \exp(i \omega t)}{\alpha+ \beta x+\cosh( \mu x) \exp(i \omega t)} \nonumber \\
& + &  \left[ \frac{\beta+ \mu \sinh( \mu x) \exp(i \omega t) }{\alpha+ \beta x +\cosh (\mu x) \exp(i \omega t) }\right]^2
\end{eqnarray} 
 which is periodic with period $T= 2 \pi / \omega$. A typical behavior of the real and imaginary parts of $V(x,t)$ over one oscillation cycle is shown in Fig.2(a). Note that, for $\alpha=\beta=0$, the potential $V(x,t)$ becomes stationary (time-independent), namely $V(x,t)=- \mu^2 \sech^2 ( \mu x)$, which belongs to the well-known class of Hermitian reflectionless potentials \cite{uff2,uff3}. To show the absence of Floquet scattering in the time-periodic case, let us construct the solution to Eq.(1), with the potential $V(x,t)$ given by Eq.(12), corresponding to a free particle of momentum $p$ and energy $E=p^2/2$ incident on the left side, i.e. of the form defined by Eqs.(3-6). Such a solution is simply given by the Daboux transformation of the free-particle state $\phi(x,t) \propto \exp(ipx-iEt)$, i.e. [see Eq.(10)]
 \begin{eqnarray}
 \psi(x,t) & = & \frac{1}{ip+\mu} \left( ip- \frac{\beta + \mu \sinh ( \mu x) \exp(i \omega t) }{\alpha +\beta x + \cosh (\mu x) \exp(i \omega t) }\right) \times \nonumber \\
 & \times & \exp(ipx-iEt)
 \end{eqnarray}
Note that, since
 \begin{eqnarray}
 \psi(x,t)  \sim \exp(ipx-iEt)
 \end{eqnarray}
for $x \rightarrow -\infty$, and
 \begin{eqnarray}
 \psi(x,t)  \sim \frac{ip- \mu}{ip + \mu} \exp(ipx-iEt)
 \end{eqnarray}
for $ x \rightarrow + \infty$, a comparison of Eqs.(14) and (15) with Eqs.(3-6) yields the following expression for the reflection and transmission coefficients of the various Floquet channels for left-side particle incidence
\begin{eqnarray}
r_n^{(left)}(E) & = & 0 , \; \; t_n^{(left)}(E)=0 \; {\rm for} \; n \neq 0,  \nonumber \\
t_0^{(left)}(E) & = & \frac{ip-\mu}{ip+\mu}
\end{eqnarray}
with $p=\sqrt{2E}$. Equations (16) prove that in the oscillating potential (12) the Floquet scattering is fully suppressed. Note that $|t_0^{(left)}(E)|=1$, i.e. the scattering matrix is unitary
in spite the potential is non-Hermitian. Note also that the phase of the transmission amplitude $t_0^{(left)}(E)$ does not depend on the frequency $\omega$ of the oscillating potential, and it is equal to the one of the static reflectionless potential obtained in the limit $\alpha =\beta=0$ (see, for instance, \cite{uff2}). The effect of this phase term is to slightly {\it advance} and distort (narrow) a  wave packet that crosses the potential well, like in the Hermitian case \cite{uff3}. The group delay (phase time) is given by
\begin{equation}
\Delta \tau=\frac{1}{p}{\rm Im} \left\{ \frac{\partial}{\partial p} {\log} (t_0^{(left)}) \right\}=-\frac{2 \mu}{p (\mu^2+p^2)}
\end{equation}
which is negative, indicating an advancement of the wave packet.
The previous relations (16) and (17), derived for left-side particle incidence, are also valid for a particle incident from the right side of the oscillating potential well; this result can be readily proven by considering the asymptotic behavior of the Darboux transformation $\psi(x,t)= \hat{\mathcal{D}} \phi (x,t)$ of the free-particle plane wave $\phi(x,t)=\exp(-ipx-iEt)$, describing a freely moving particle moving from the right to the left sides with momentum $p$. Hence, like in the Hermtian case our non-Hermitian potential well behaves symmetrically for left and right-side incidence, contrary to e.g. $\mathcal{PT}$-symmetric potential wells or barriers which show left/right handedness \cite{NH1}. In fact, while in a $\mathcal{PT}$-symmetric potential the imaginary part of the potential is an odd function of space, in our case it is an even function for spatial inversion $x \rightarrow -x$, see e.g. Fig.2(a).

 \par
The second family of oscillating quantum wells without Floquet scattering is synthesized by assuming the following solution  $u(x,t)$ to the equation $( i \partial_t-\hat{\mathcal{H}_0}) u=0$
 \begin{equation}
 u(x,t)=i \alpha+ \beta x+ \cos(\mu x) \exp(-i \omega t)
 \end{equation}
 where $\mu>0$ is an arbitrary real-valued parameter, $\omega=\mu^2 / 2$, and $\alpha$, $\beta$ are two arbitrary real-valued parameters with $|\alpha|>1$. Such a condition ensures that $u(x,t)$ is a nonvanishing function, and thus the partner potential $V(x,t)$ is not singular.  The explicit form of the potential $V(x)$, obtained from Eqs.(9) and (18) with $L(t)=1$ and $V_0(x,t)=0$, reads
 \begin{eqnarray}
 V(x,t) & = & \frac{\mu^2 \cos (\mu x) \exp(- i \omega t))}{i \alpha + \beta x + \cos ( \mu x) \exp(-i \omega t)} \nonumber \\
 & + & \left[ \frac{\beta - \mu \sin ( \mu x) \exp(- i \omega t)}{i \alpha + \beta x + \cos (\mu x) \exp(-i \omega t)} \right]^2
 \end{eqnarray}
 A typical behavior of the real and imaginary parts of the potential $V(x,t)$, defined by Eq.(19), is shown in Fig.2(b). As in the previous case, the solution to Eq.(1), with the potential $V(x,t)$ given by Eq.(19) and corresponding to a free particle of momentum $p$ and energy $E=p^2/2$ incident on the left side, is obtained from the Daboux transformation of the free-particle state $\phi(x,t)=\exp(ipx-iEt)$ and reads
 \begin{eqnarray}
 \psi(x,t) & = & \left( 1- \frac{1}{ip} \frac{\beta - \mu \sin ( \mu x) \exp(-i \omega t) }{i \alpha +\beta x + \cos (\mu x) \exp(-i \omega t) }\right) \times \nonumber \\
 & \times & \exp(ipx-iEt)
 \end{eqnarray}
Note that, since
 \begin{eqnarray}
 \psi(x,t)  \sim \exp(ipx-iEt)
 \end{eqnarray}
for $x \rightarrow -\infty$ and $ x \rightarrow + \infty$,  the reflection and transmission coefficients of the various Floquet channels for left-side incidence are merely given by
\begin{equation}
r_n^{(left)}(E)=0 , \; \; t_n^{(left)}(E)=0 \; {\rm for} \; n \neq 0, \; \; t_0^{(left)}(E)=1
\end{equation}
with $p=\sqrt{2E}$. Equations (22) show that in the oscillating potential (19) the Floquet scattering is fully suppressed. In addition, since $t_0^{(left)}(E)=1$, the oscillating well is invisible, i.e. it does not introduce any delay/advancement nor distortion of a wave packet crossing the potential.  The same result holds for a particle incident onto the  oscillating potential well from the right side.
 
\begin{figure}[t]
\includegraphics[width=8cm]{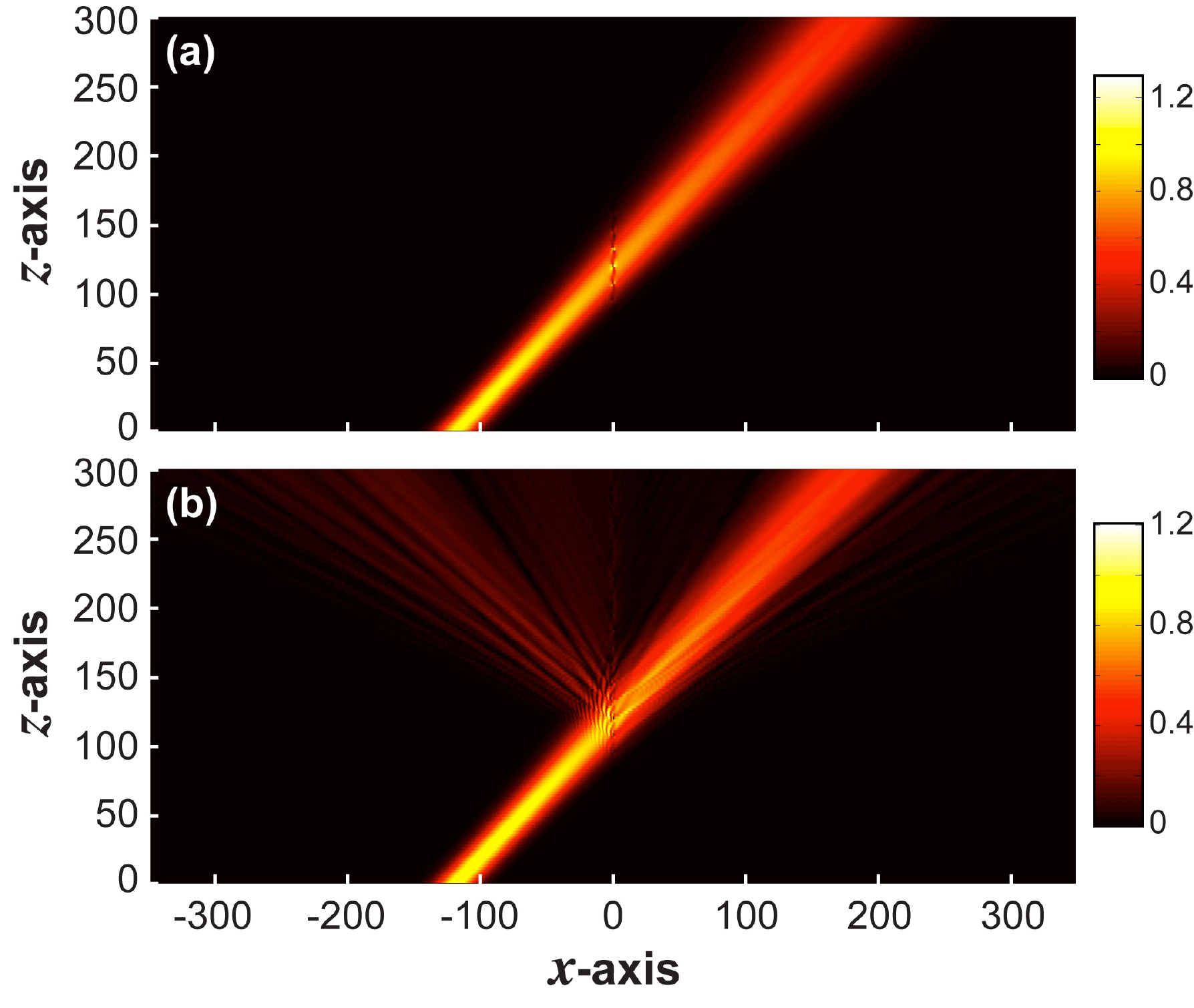}
\caption{(Color online). Evolution of a Gaussian wave packet (snapshot of $|\psi(x,t)|$) in (a) the oscillating complex potential well $V(x,t)$ defined by Eq.(12), and (b) in the associated Hermitian potential well $V_{Herm}(x,t)={\rm Re} \{ V(x,t) \}$. Parameter values are given in the text.}
\end{figure}

\begin{figure}[t]
\includegraphics[width=8cm]{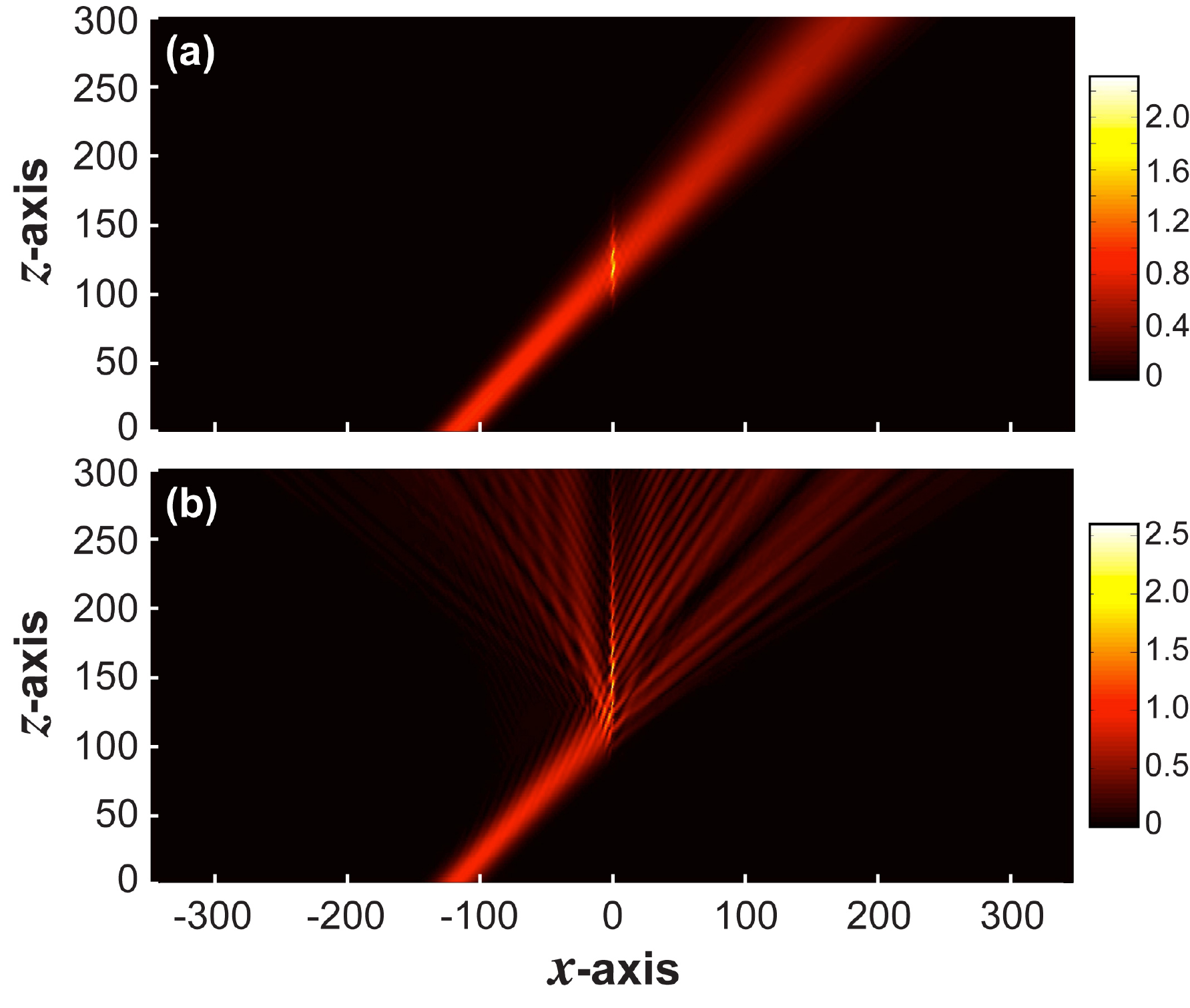}
\caption{(Color online). Same as Fig.3, but for the potential defined by Eq.(19). Parameter values are given in the text.}
\end{figure}

\section{Wave packet propagation: numerical results}
We have checked the absence of Floquet scattering in the two families of oscillating potential wells, introduced in the previous section and defined by Eqs.(12) and (19), by direct numerical simulations of the time-dependent Schr\"{o}dinger equation 
 using an accurate pseudospectral split-step method.  As an initial condition at time $t=0$, we assumed a Gaussian wave packet, of width $w$ and localized at the position $x_0<0$ far from the potential well, with a mean momentum $p$, i.e. $ \psi(x,0)=\exp[-(x-x_0)^2/w^2] \exp(ipx)$. Figure 3(a) shows the evolution of the amplitude probability $|\psi(x,t)|$ of the wave packet crossing the complex oscillating potential (12) for parameter values $\alpha=0.9$, $\beta=0$, $\mu=1$, $w=15$,  $p=1$ and $x_0=-120$. The figure clearly shows the absence of Floquet scattering. For comparison, in Fig.3(b) we show the evolution of the amplitude probability $|\psi(x,t)|$ in the oscillating Hermitian potential well $V_{Herm}(x)={\rm Re} \{ V(x) \}$, obtained by taking the real part solely of $V(x)$. In this case the wave packet clearly undergoes Floquet scattering when crossing the potential well. A similar behavior is found by considering wave packet scattering from the complex oscillating potential defined by Eq.(19), which is shown in Fig. 4(a) for parameter values $\alpha=2$, $\beta=2$, $\mu=1$, $w=15$,  $p=1$ and $x_0=-120$. Figure 4(b) shows, for comparison, the wave packet evolution scattered off  by the Hermitian potential $V_{Herm}(x)={\rm Re} \{ V(x) \}$, obtained by taking the real part solely of $V(x)$. According to the analysis of Sec.III, the oscillating potential (12) does not scatter the wave packet, however it introduces an advancement of the wave packet, given by Eq.(17). For parameter values used in the simulations, the group delay is estimated to be $\Delta \tau= - 1$, which corresponds to a spatial advancement of the wave packet, for a given time $t$, of $\Delta x=p|\Delta \tau|= 1$. Note that such an advancement is rather small as compared to the width $2w=30$ of the initial wave packet. Conversely, the oscillating potential (19) is fully invisible, i.e. no delay/advancement nor distortion of the wave packet is introduced after crossing the potential well. This is clearly shown in Fig.5, which depicts the spatial profile of $|\psi(x,t)|^2$ at time $t=300$, after crossing the well, for the freely moving wave packet  (i.e., in the absence of the potential, thin solid line), for the wave packet crossing the potential (12) (dotted curve), and for the wave packet crossing the potential (19) (dashed line). The three curves are almost overlapped. However, a closer inspection of the wave packet tails, shown in the inset of Fig.5, indicates that the potential (12) introduces a small advancement, whereas the potential (19) does not, in agreement with the theoretical analysis.  The numerical simulations shown in Figs.3-5 refer to a wave packet propagating from the left to the right side, however the same reults are obtained by considering the scattering dynamics for a wave packet that crosses the oscillating potential well from the right to the left sides. 

\begin{figure}[t]
\includegraphics[width=8cm]{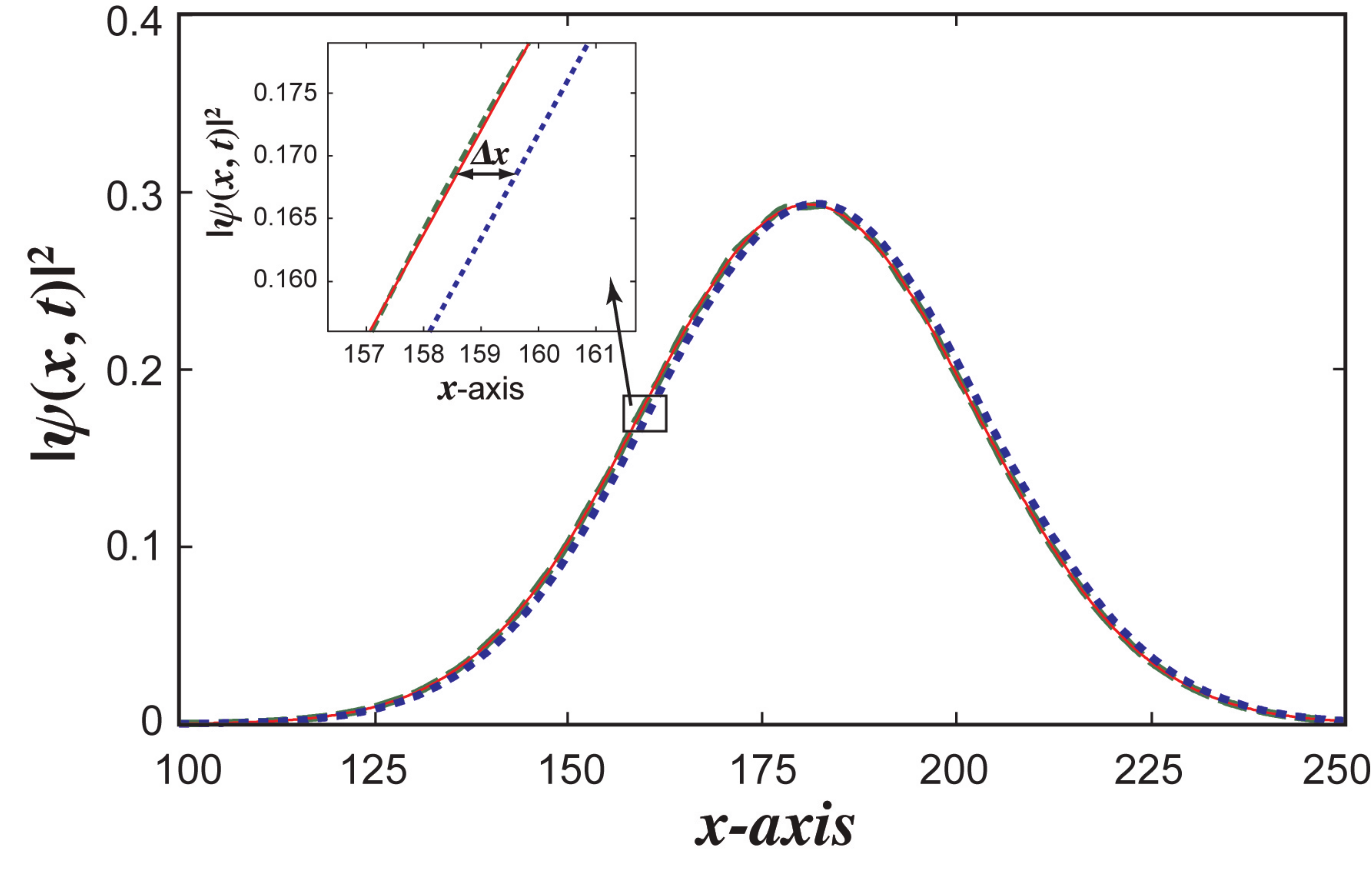}
\caption{(Color online). Behavior of the wave packet distribution probability $|\psi(x,t)|^2$ at time $t=300$, after crossing the scattering potential, corresponding to the free particle (thin solid curve), the potential (12) (dotted curve), and the potential (19) (dashed curve). The three distributions are almost overlapped. The inset depicts an enlargement of the distribution tails, clearly showing the advancement $\Delta x$ of the wave packet crossing the potential (12).}
\end{figure}

\section{Conclusions and discussion}
A quantum particle crossing a potential barrier or well, oscillating in time at frequency $\omega$, undergoes rather generally  to Floquet scattering, with the particle being transmitted or reflected with an energy $E'=E \pm n \omega$ which can differ from the energy $E$ of the incident particle owing to the absorption or emission of $n$ quanta from the photon field.  Floquet scattering is of fundamental importance in different areas of physics and plays a major role in quantum transport properties of ac-driven microscopic and mesoscopic
systems. It is also analogous to the problem of Bragg scattering of monochromatic matter or optical waves from a diffraction grating.  On the other hand, it is well known that  for certain static potential wells, i.e. for the so-called reflectionless potentials,  scattering can be absent for any energy of the incidence particle. A natural question arises whether {\it Floquet scattering} can be absent in a certain class of oscillating potentials, an issue which has not been investigated yet. In this work we have shown that  families of non-Hermitian oscillating potential wells can be synthesized by application of the Darboux transformation to the time-dependent Schr\"{o}dinger equation. While with this technique we could not synthesize scattering-free oscillating potentials of Hermitian type, by allowing the potential to become-complex-valued, i.e. the Hamiltonian non-Hermitian, two families of scattering-free and norm-conserving oscillating potentials have been introduced, admitting the analytical expressions defined by Eqs.(12) and (19). The latter family of oscillating wells, in addition to suppressing Floquet scattering, is also fully invisible. We note that, as compared to unidirectional suppression of Bragg scattering recently predicted and observed in the context of $\mathcal{PT}$-symmetric non-Hermitian Hamiltonians \cite{cazz2,Lin,Longhi10}, in our complex potentials the absence of Floquet scattering is bidirectional, i.e. it occurs for both left and right side particle incidence.  Our results provide new physical insights into the basic problem of scattering from oscillating potentials, and are expected to stimulate further theoretical and experimental studies. On the experimental side, the realization of complex potentials with tailored profiles, as predicted by Eqs.(12) and (19),  remains a challenging task. Synthetic optical structures with controlled gain and loss regions might be  considered as potentially-accessible systems to test Floquet scattering in non-Hermitian potentials \cite{cazz3}, however the search for simpler potentials could facilitate the experimental realization of scattering-free  potential. In particular, it remains open the question whether Floquet scattering can be absent in {\it Hermitian} oscillating potential wells, which could be of interest in problems of quantum transport in mesoscopic systems. 
 
\acknowledgments

This work was supported by the Fondazione Cariplo (Grant No. 2011-0338).


\begin{thebibliography}{99}



\bibitem{Butt1}
M. B\"{u}ttiker and R. Landauer, Phys. Rev. Lett. {\bf 49}, 1739
(1982).

\bibitem{Butt2}
R. Landauer and Th. Martin, Rev. Mod. Phys. {\bf 66}, 217 (1994).

\bibitem{Butt3}
V. Gasparian, Superlattices and Microstructures {\bf 23},  809  (1998).


\bibitem{chaos1}
J. L. Mateos and J. V. Jose, Physica A {\bf 257}, 434 (1998).

\bibitem{chaos2}
M. Henseler, T. Dittrich, and K. Richter, Europhys. Lett. {\bf 49}, 289 (2000).


\bibitem{chaos3}
E. D. Leonel and P. V. E. McClintock, Phys. Rev. E
{\bf 70}, 016214 (2004).

\bibitem{chaos4}
F.R.N. Koch, F. Lenz, C. Petri, F.K. Diakonos, and P. Schmelcher, Phys. Rev. E {\bf 78}, 056204 (2008).

\bibitem{book}
{\it Dynamical Tunneling: Theory and Experiment}, edited by S. Keshavamurthy and P. Schlagheck 
(Taylor \& Francis, Boca-Raton, 2011). 

\bibitem{r7}
M. Henseler, T. Dittrich, and K. Richter, Phys. Rev. E
{\bf 64}, 046218 (2001).

\bibitem{r8}
W. A. Lin and L. E. Ballentine, Phys. Rev. Lett. {\bf 65}, 2927
(1990).

\bibitem{Platero}
G. Platero and R. Aguado, Phys. Rep. {\bf 395}, 1 (2004). 

\bibitem{Hanggi}
S. Kohler, J. Lehmann, and P. H\"{a}nggi, Phys. Rep. {\bf 406}, 379 (2005).

\bibitem{tras1}
W. Cai, T. F. Zheng, P. Hu, B. Yudanin, and M. Lax, Phys. Rev. Lett. {\bf 63}, 418 (1989). 

\bibitem{tras2}
W.-T. Lu, S.-J. Wang, W. Li, Y.-L. Wang, C.-Z. Ye, and H. Jiang, J. Appl. Phys. {\bf 111}, 103717 (2012). 

\bibitem{tras3}
M. A. Zeb, K. Sabeeh, and M. Tahir,
Phys. Rev. B 78, 165420 (2008)  

\bibitem{meso}
C.-H. Yan and L.-F. Wei, J. Phys.: Condens. Matter {\bf 22}, 185301 (2010).

\bibitem{Flo}
W. Li and L. E. Reichl, Phys. Rev. B {\bf 60}, 15732 (1999). 

\bibitem{r3} A. Pimpale, S. Holloway, and R. J. Smith, J. Phys. A {\bf 24},
3533 (1991).

\bibitem{r4}
V. A. Fedirko and V. V. Vyurkov, Phys. Status Solidi B
{\bf 221}, 447 (2000).

\bibitem{r5c}
M. Moskalets and M. B\"{u}ttiker, Phys. Rev. B {\bf 66}, 205320 (2002).

\bibitem{r5b}
D. Shin and J. Hong, Phys. Rev. B {\bf 70}, 073301 (2004).

\bibitem{r5}
M. Garttner, F. Lenz, C. Petri, F. K. Diakonos and P.
Schmelcher, Phys. Rev. E {\bf 81}, 051136 (2010).

\bibitem{r5graphene}
R. Zhu and H. Chen, Appl. Phys. Lett. {\bf 95}, 122111 (2009). 

\bibitem{r6}
A. Emmanouilidou and L. E. Reichl, Phys. Rev. A {\bf 65},
033405 (2002).

\bibitem{r9}
L. M. Pecora, H. Lee, D. H. Wu, T. Antonsen, M. J. Lee,
E. Ott, Phys. Rev. E {\bf 83}, 065201 (2011).

\bibitem{r10} 
D. A. Steck, W. H. Oskay, and M. G. Raizen, Science
{\bf 293}, 274 (2001).

\bibitem{r11}
D. A. Steck, W. H. Oskay, and M. G. Raizen, Phys. Rev.
Lett. {\bf 88}, 120406 (2002).

\bibitem{r12}
F. Grossmann, T. Dittrich, P. Jung, and P. H\"{a}nggi, Phys.
Rev. Lett. {\bf 67}, 516 (1991).

\bibitem{Grifoni}
M. Grifoni and P. H\"{a}nggi, Phys. Rep. {\bf 304},  229 (1998).
 
\bibitem{r13}
S. Rahav and P. W. Brouwer, Phys. Rev. B {\bf 74}, 205327
(2006).

\bibitem{Vor}
I. Vorobeichik, R. Lefebvre, and N. Moiseyev,
Europhys. Lett. {\bf 41}, 111 (1998).

\bibitem{quantum}
G. Sulyok, J. Summhammer, and H. Rauch, Phys. Rev. A {\bf 86}, 012124 (2012).

\bibitem{Berry1}
M.V. Berry, J. Phys. A {\bf 31}, 3493 (1998).

\bibitem{Berry2}
M. V. Berry and D.H.J. O'Dell, J. Phys. A {\bf 31}, 2093 (1998).

\bibitem{Longhi}
S. Longhi, Opt. Lett. {\bf 30}, 2781 (2005).

\bibitem{Moi}
N. Moiseyev, {\it 
Non-Hermitian Quantum Mechanics} (Cambridge University Press, London,  Cambridge, 2011).

\bibitem{Bender}
C. M. Bender, Rep. Prog. Phys. {\bf 70}, 957 (2007).

\bibitem{Keller}
C. Keller, M.K. Oberthaler, R. Abfalterer, S. Bernet, J. Schmiedmayer, and A. Zeilinger, Phys. Rev. Lett. {\bf 79}, 3327 (1997).

\bibitem{Ober}
R. St\"{u}tzle, M.C. G\"{o}bel, Th. H\"{o}rner, E. Kierig, I. Mourachko, M.K. Oberthaler, M. A. Efremov, M.V. Fedorov, V. P. Yakovlev, K. A. H. van Leeuwen, and W. P. Schleich, Phys. Rev. Lett. {\bf 95}, 110405 (2005).


\bibitem{cazz1}
C. E. R\"uter, K. G. Makris, R. El-Ganainy, D. N. Christodoulides, M. Segev and D. Kip, Nature Phys. {\bf 6}, 192 (2010).

\bibitem{cazz2}
A. Regensburger, C. Bersch, M.-A. Miri, G. Onishchukov, D. N. Christodoulides, and U. Peschel, Nature {\bf 488}, 167 (2012).

\bibitem{cazz3}
L. Feng, Y.-L. Xu, W. S. Fegadolli, M.-H. Lu, J. E. B.
Oliveira, V. R. Almeida, Y.-F. Chen, and A. Scherer,
Nature Mater. {\bf 12}, 108 (2013).

\bibitem{Lin}
Z. Lin, H. Ramezani, T. Eichelkraut, T. Kottos, H. Cao, and D. N. Christodoulides, Phys. Rev. Lett. {\bf 106}, 213901 (2011).

\bibitem{Longhi10}
S. Longhi,
Phys. Rev. A {\bf 81}, 022102  (2010). 

\bibitem{Longhi11}
S. Longhi, J. Phys. A {\bf 44}, 485302 (2011).

\bibitem{Jones}
E.-M. Graefe and H. F. Jones, 
Phys. Rev. A {\bf 84}, 013818 (2011). 

\bibitem{NH1}
Z. Ahmed, Phys. Rev. A {\bf 64},  042716 (2001).

\bibitem{NH2}
M. Znojil, Phys. Lett. A {\bf 285}, 7 (2001). 

\bibitem{NH3}
Z. Ahmed,  Phys. Lett. A {\bf 324}, 152 (2004).

\bibitem{NH4}
F. Cannata, J.-P. Dedonder, and A. Ventura,
Annals of Phys.  {\bf 322}, 397 (2007).

\bibitem{NH5}
G. Levai, P. Siegl, and M. Znojil, J. Phys. A {\bf 42}, 295201 (2009).

\bibitem{NH6}
Z. Ahmed, J. Phys. A {\bf 45},  032004  (2012). 

\bibitem{note1}
In the Hermitian case, the scattering  process conserves the quasi-energy, however the energy of the transmitted (or reflected) particle is generally not conserved owing to energy exchange with the photons of the driving field. 

\bibitem{note2}
Partial, but not complete, absence of grating diffraction orders was previously predicted to occur by Berry (Refs.\cite{Berry1,Berry2}) for normal incidence in a  complex sinusoidal grating potential. In such a $\mathcal{PT}$-symmetric potential, unidirectional Bragg scattering  was shown to occur in Refs.\cite{Lin,Longhi10,Longhi11,Jones}.

\bibitem{uff1}
I. Kay and H.E. Moses, J. Appl. Phys. {\bf 27}, 1503 (1956) 

\bibitem{uff2}
A. K. Grant and J.L. Rosner, J. Math. Phys. {\bf 35}, 2142 (1994).

\bibitem{uff3}
J. Lekner, Am. J. Phys. {\bf 75}, 1151 (2007).

\bibitem{uff4}
S. P. Maydanyuk, Annals of Phys. {\bf 316}, 440 (2005).

\bibitem{D1}
B.F. Samsonov and L. A. Shekoyan, Phys. Atom. Nucl. {\bf 63}, 657 (2000).

\bibitem{D2}
B.F. Samsonov, M. L. Glasser, and L. M. Nieto, J. Phys. A {\bf 36}, L585 (2003).

\bibitem{D3}
B.F. Samsonov, J. Phys. A {\bf 37},  10273 (2004).

\bibitem{D4}
A.A. Suzko and A. Schulze-Halberg, J. Phys. A {\bf 42}, 295203 (2009).












\end{thebibliography}
\end{document}